\documentclass[prd,nofootinbib,showpacs,preprintnumbers,amsmath,amssymb]{revtex4}

\usepackage{graphicx}
\usepackage{dcolumn}
\usepackage{bm}

\def\bear{\begin{eqnarray}}
\def\ear{\end{eqnarray}}

\begin{document}

\begin{flushright}
YITP-13-63
\end{flushright}

\title{Second Quantized Scalar QED in Homogeneous Time-Dependent Electromagnetic Fields}

\author{Sang Pyo Kim}\email{sangkim@kunsan.ac.kr}
\affiliation{Department of Physics, Kunsan National University, Kunsan 573-701, Korea}
\affiliation{Yukawa Institute for Theoretical Physics, Kyoto University, Kyoto 606-8502, Japan}
\affiliation{Center for Relativistic Laser Science, Institute for Basic Science, Gwangju 500-712, Korea}

\medskip

\date{\today}

\begin{abstract}
We formulate the second quantization of a charged scalar field in homogeneous, time-dependent electromagnetic fields, in which the Hamiltonian is an infinite system of decoupled, time-dependent oscillators for electric fields, but it is another infinite system of coupled, time-dependent oscillators for magnetic fields. We then employ the quantum invariant method to find various quantum states for the charged field. For time-dependent electric fields, a pair of quantum invariant operators for each oscillator with the given momentum plays the role of the time-dependent annihilation and the creation operators, constructs the exact quantum states, and gives the vacuum persistence amplitude as well as the pair-production rate. We also find the quantum invariants for the coupled oscillators for the charged field in time-dependent magnetic fields and advance a perturbation method when the
magnetic fields change adiabatically. Finally, the quantum state and the pair production are discussed when a time-dependent electric field is present in parallel to the magnetic field.
\end{abstract}

\maketitle

\section{Introduction} \label{sec1}

Recent development of intense laser sources has brought intensive study of quantum motions of charged particles in electromagnetic fields \cite{DMHK12}. In astrophysics, magnetars, strongly magnetized neutron stars, have magnetic fields over the critical strength \cite{harding-lai06}. The quantum states of charged particles in such strong electromagnetic fields provide an essential ingredient to understand the vacuum structure and the pair production. The interaction of virtual pairs of the Dirac sea with a strong external electromagnetic field polarizes the vacuum and thus leads to a nonlinear effective action beyond the Maxwell theory. In particular, the complex effective action in an electric field implies the vacuum instability due to the pair production of charged particles. Hence, intense lasers have been proposed as a feasible tool to probe the vacuum structure via the pair production and the vacuum polarization (for review and references, see Refs. \cite{dunne09,gies09}).

 In this paper, we formulate the second quantized scalar field in homogeneous, time-dependent electromagnetic fields and employ the invariant operator method to find the quantum states of charged scalars. The Hamiltonian from the field action in a homogeneous, time-dependent electric field is a system of infinite number of decoupled, time-dependent oscillators, while it is equivalent to another system of infinite number of coupled oscillators in a time-dependent magnetic field with or without a parallel electric field. At the level of the first quantization, the Klein-Gordon (KG) equation in the vector potential separates into each Fourier mode for the time-dependent electric field \cite{KLY08}, while the KG equation in the time-dependent magnetic field does not separate into each harmonic wave function, which corresponds to an instantaneous Landau state as shown in Ref. \cite{kim14a}.
In fact, the KG equation in a time-dependent magnetic field in the two-component first-order formalism \cite{kim14a} is equivalent to the Wheeler-DeWitt equation for the Friedmann-Robertson-Walker universe minimally coupled to a massive scalar field \cite{kim92,kim13}. Interestingly, the third quantized universe coupled to a massless scalar is equivalent to a system of infinite number of decoupled oscillators with a time-dependent mass and time-dependent frequencies \cite{kim14b}, which is analogous to the charged field in a time-dependent electric field.

The functional Schr\"{o}dinger picture provides a second quantized field theory, which extends the time-dependent Schr\"{o}dinger equation for quantum mechanical systems to quantum fields \cite{FHM85,guth-pi85}. Thus, in the second quantized scalar field, the quantum theory is the functional Schr\"{o}dinger equation with time-dependent Hamiltonians for a charged scalar field in external electromagnetic fields. In the first case of time-dependent electric fields, the Gaussian wave functional can be directly found in terms of the covariance \cite{kiefer92}. In the Fourier decomposition, the functional Schr\"{o}dinger equation becomes the Schr\"{o}dinger equation for decoupled, time-dependent oscillators, and the quantum state of the field is the product of the time-dependent wave function for each oscillator. It has long been known that an oscillator with time-dependent frequency and/or mass has a quantum invariant, known as the Lewis-Riesenfeld invariant, whose eigenstate provides an exact solution of the Schr\"{o}dinger equation up to a time-dependent phase factor \cite{lewis-riesenfeld69}. Hence, in the former case of electric fields, we may employ the time-dependent annihilation and the creation operators, also quantum invariants, and construct not only excited states for each time-dependent oscillator \cite{MMT70,kim-kim99,kim-page01} but also a thermal state \cite{kim-lee00,kim-page13}.
In the second case of magnetic fields, however, the Hamiltonian for the functional Schr\"{o}dinger equation is equivalent to coupled, time-dependent oscillators, and hence, we should employ the invariants for coupled, time-dependent oscillators \cite{leach77,ji-hong98,dodonov00,lee01}.

The quantum invariant method for time-dependent oscillators is very useful in constructing various quantum states from the vacuum state, ranging from excited states to coherent states and even to thermal states (for review and references, see Refs. \cite{dodonov-manko89,lohe09}). By analogy with time-independent oscillators, the time-dependent annihilation and the creation operators for time-dependent oscillators, quantum invariants linear in momentum and position
operators, may be used to find the Fock space of all the number states, which are the exact solutions for the time-dependent Schr\"{o}dinger equation \cite{MMT70,kim-lee00}. Furthermore, it can be readily used in constructing thermal states and coherent thermal states, which are also the exact quantum states
\cite{kim-lee00,kim-page13}. In scalar quantum electrodynamics (QED) in time-dependent electric fields, the time-dependent vacuum state leads to the pair-production rate \cite{kim-schubert11,KLR12} and also to the in-out scattering matrix between the remote past and the remote future, which yields the renormalized one-loop effective action \cite{KLY08}. The quantum invariant method can also be used to describe the quantum motion of charged particles in strong electromagnetic fields. In the case of the adiabatically changing magnetic fields, the invariants may be found for coupled, time-dependent oscillators by treating the off-diagonal Hamiltonian as a perturbation for the diagonal Hamiltonian, decoupled time-dependent oscillators. We advance a time-dependent perturbation theory to find the improved states. The method is valid for the time-dependent magnetic fields parallel to electric fields.

The organization of this paper is as follows. In Sec. \ref{sec2}, in scalar QED we formulate the second quantized scalar field in homogeneous, time-dependent electric fields or magnetic fields prescribed by vector potentials. The Hamiltonian from the field action in time-dependent electric fields is decomposed by the Fourier modes and is expressed as a sum of decoupled, time-dependent oscillators. In time-dependent magnetic fields, the Hamiltonian is decomposed both by the Fourier modes along the longitudinal direction and by the Landau states in the transverse plane that diagonalize the transverse part of the Hamiltonian. In Sec. \ref{sec3}, we study the time-dependent Schr\"{o}dinger equation for the quadratic Hamiltonian in the extended phase space of positions and momenta and search for the quantum invariants for the field. In Sec. \ref{sec4}, we find the time-dependent annihilation and the creation operators, construct the Fock space for the charged scalar field in time-dependent electric fields, and discuss the pair production and the vacuum persistence amplitude. In Sec. \ref{sec5}, we extend the quantum invariant method to find the time-dependent annihilation and the creation operators for the charged field in time-dependent magnetic fields. We advance a perturbation method to find the improved quantum states beyond the Landau states when the magnetic fields change adiabatically. We show that the Hamiltonian has the same algebraic structure even when an electric field is added in parallel to the magnetic field and, thus, the same form of invariants. In Sec. \ref{sec6}, we discuss the physical implications of the result of this paper.

\section{Second Quantized Scalar Field} \label{sec2}

In scalar QED, we formulate the second quantization of a spinless, charged  scalar field in homogeneous, time-dependent electromagnetic fields, first by expressing the Hamiltonian from the field action through a spectral method, and then by finding quantum states via the functional Schr\"{o}inger equation. For that purpose, we consider the action for a spinless scalar with charge $q$ and mass $m$ [in units of $\hbar = c =1$ and the spacetime signature $\eta^{\mu \nu} = (+, -, -, -)$]
\begin{eqnarray}
S = \int dt d^3 {\bf x} \bigl[ \eta^{\mu \nu} (\partial_{\mu} + iqA_{\mu} ) \phi^{*} (\partial_{\mu} - iqA_{\mu} ) \phi - m^2 \phi^{*} \phi \bigr]
\end{eqnarray}
in an electromagnetic field, which is given by the four-vector as
\begin{eqnarray}
{\bf E} = \nabla A_0 - \frac{\partial {\bf A}}{\partial t}, \quad {\bf B} = \nabla \times {\bf A}.
\end{eqnarray}
Introducing the conjugate momenta for each $\phi$ and $\phi^*$
\begin{eqnarray}
\pi = \frac{\partial {\cal L}}{\partial \dot{\phi}} = \dot{\phi}^{*} + iqA^{0} \phi^{*}, \quad
\pi^{*} = \frac{\partial {\cal L}}{\partial \dot{\phi}^{*}} = \dot{\phi} - iqA^{0}\phi,
\end{eqnarray}
we obtain the Hamiltonian for the field
\begin{eqnarray}
H(t) = \int d^3 {\bf x} \Bigl[ \pi^{*} \pi + i qA^{0} (\pi \phi - \pi^{*} \phi) + (\partial_{k} + i qA^{*}_{k}) \phi^{*} (\partial_{k} - i qA_{k}) \phi
+ m^2 \phi^{*} \phi \Bigr].
\end{eqnarray}
Then, the quantum dynamics is governed by the functional Schr\"{o}dinger equation
\begin{eqnarray}
i \frac{\partial}{\partial t} \Psi (t, \phi, \phi^{*}) = \hat{H} (t) \Psi (t, \phi, \phi^{*}). \label{sch eq}
\end{eqnarray}
In this paper, we consider only the vector potential for a homogeneous, time-dependent electric and magnetic field along a fixed direction of the form
\begin{eqnarray}
{\bf A}_{\parallel} (t) = - \int^t_{- \infty}  dt' {\bf E}_{\parallel} (t'),  \quad {\bf A}_{\perp} (t, {\bf x}_{\perp})  = \frac{1}{2} {\bf B} (t) \times {\bf x}. \label{vec pot}
\end{eqnarray}
We further assume that ${\bf E}_{\parallel} (-\infty) = 0$ and ${\bf B} (- \infty) = {\bf B}_0$ so that the initial state is either the Minkowski vacuum or the Landau state, respectively.

Firstly, in the case of homogeneous, time-dependent electric fields, we decompose the fields by the Fourier modes as
\begin{eqnarray}
\phi (t, {\bf x}) = \int \frac{d^3 {\bf k}}{(2 \pi)^{3}} \phi_{\bf k} (t)  e^{i {\bf k} \cdot {\bf x}}, \quad \phi^* (t, {\bf x}) = \int \frac{d^3 {\bf k}}{(2 \pi)^{3}} \phi^*_{\bf k} (t)  e^{-i {\bf k} \cdot {\bf x}}, \label{fourier}
\end{eqnarray}
and obtain the time-dependent Hamiltonian
\begin{eqnarray}
H(t) = \int \frac{d^3 {\bf k}}{(2 \pi)^{3}} \Bigl[ \pi_{\bf k}^{*} \pi_{\bf k} + \omega_{\bf k}^2 (t) \phi_{\bf k}^{*} \phi_{\bf k} \Bigr], \label{E-ham}
\end{eqnarray}
where $\pi_{\bf k} = \dot{\phi}^*_{\bf k}$ and $\pi_{\bf k}^* = \dot{\phi}_{\bf k}$ are canonical momenta, and the time-dependent frequencies are
\begin{eqnarray}
\omega_{\bf k}^2 (t) = (k_{\parallel} - q A_{\parallel} (t))^2 + {\bf k}_{\perp}^2 + m^2.
\end{eqnarray}
The classical Heisenberg equation
\begin{eqnarray}
 \ddot{\phi}_{\bf k} (t) + \omega_{\bf k}^2 (t) \phi_{\bf k} (t) =0, \label{E-eq}
\end{eqnarray}
is the corresponding Fourier mode of the KG equation, which explains the equivalence between the quantum invariant method in this paper and the conventional canonical quantum field theory. As mentioned in Introduction, the quantum invariant method has the merit of diversity of quantum states.

Secondly, in the case of homogeneous, time-dependent magnetic fields along the $z$-direction, after Fourier-decomposing the fields along the longitudinal direction, we find the Lagrangian
\begin{eqnarray}
L(t) = \int d^2 {\bf x}_{\perp} \Bigl[ \dot{\phi}^{*}_{k_z} ({\bf x}_{\perp}) \dot{\phi}_{k_z} ({\bf x}_{\perp}) - \phi^{*}_{k_z} ({\bf x}_{\perp}) \Bigl(H_{\perp} (t)+ k_z^2 + m^2 \Bigr) \phi_{k_z} ({\bf x}_{\perp}) \Bigr], \label{B-lag}
\end{eqnarray}
where
\begin{eqnarray}
H_{\perp} (t) = {\bf p}_{\perp}^2 + \Bigl(\frac{qB(t)}{2} \Bigr)^2 {\bf x}_{\perp}^2 - qB(t) L_z \label{tr-ham}
\end{eqnarray}
is the Hamiltonian transverse to the magnetic field, and $L_z$ is the angular momentum from the orbital motion of the charge.
Following Ref. \cite{kim14a}, we diagonalize the Hamiltonian (\ref{tr-ham}) as
\begin{eqnarray}
\hat{H}_{\perp} (t) = q B(t) \bigl[2 \hat{c}_{-}^{\dagger} (t) \hat{c}_{-} (t) +1 \bigr], \label{lan ham}
\end{eqnarray}
where the time-dependent annihilation and the creation operators
\begin{eqnarray}
\hat{c}_{-} (t) = \frac{1}{\sqrt{2}} \bigl( \hat{a}_{x} (t) - i \hat{a}_{y} (t)\bigr), \quad \hat{c}_{-}^{\dagger} (t) = \frac{1}{\sqrt{2}} \bigl( \hat{a}^{\dagger}_{x} (t) + i \hat{a}^{\dagger}_{y} (t)\bigr)
\end{eqnarray}
are constructed by the annihilation operators $\hat{a}_{x} (t)$ and $\hat{a}_{y} (t)$ for the $x$-component and the $y$-component of the oscillator
\begin{eqnarray}
H_{\perp 0} (t) = {\bf p}_{\perp}^2 + \Bigl(\frac{qB(t)}{2} \Bigr)^2 {\bf x}_{\perp}^2.
\end{eqnarray}
The number states $\vert n, t \rangle$ in the oscillator representation (\ref{lan ham}), which are the instantaneous Landau states, constitute a basis. Thus, the rate of the change of each number state can be expressed by themselves, which has been first derived in Ref. \cite{kim92}, and then Eqs. (12) and (13) of Ref. \cite{kim14a} and  Eq. (13) of Ref. \cite{kim13}, which reads as
\begin{eqnarray}
\langle m, t \vert (\frac{\partial}{\partial t} \vert n, t \rangle) := \Omega_{mn} (t) = \frac{\dot{B}}{4 B} \Bigl[ \sqrt{n (n-1)} \delta_{m, n-2}
- \sqrt{(n+1) (n+2)} \delta_{m, n+2} \Bigr]. \label{ll tr}
\end{eqnarray}
Note that the coupling matrix is antisymmetric, $\Omega^T = - \Omega$.

Thus, omitting the longitudinal momentum for simplicity, we have the expansions
\begin{eqnarray}
\phi_{k_z} ({\bf x}_{\perp}, t) &=& \sum_{n} \phi_{n} (t) \Phi_n ({\bf x}_{\perp}, t), \nonumber\\
\dot{\phi}_{k_z} ({\bf x}_{\perp}, t) &=& \sum_{n} \dot{\phi}_{n} (t) \Phi_n ({\bf x}_{\perp}, t) + \sum_{n l} \phi_{n} (t) \Omega_{nl} (t) \Phi_l ({\bf x}_{\perp}, t), \label{lan exp}
\end{eqnarray}
where $ \Phi_n ({\bf x}_{\perp}, t) = \langle {\bf x}_{\perp} \vert n, t \rangle$ is the wave function for the $n$-th Landau level.
Finally, we obtain the Hamiltonian of the form
\begin{eqnarray}
H(t) = \int \frac{dk_{z}}{2\pi} \Bigl[ \sum_{n} \bigl( \pi_{n}^* \pi_{n} + \omega_n^2 \phi_n^* \phi_n \bigr) + \sum_{m n} \bigl( \pi_{m}^* \Omega_{mn}  \phi_{n}^* + \pi_{m} \Omega_{mn} \phi_{n}  \bigr) \Bigr], \label{B-ham}
\end{eqnarray}
where the canonical momenta are
\begin{eqnarray}
\pi_n &=& \dot{\phi}_n^* + \sum_{m} \phi_{m}^* \Omega_{mn} = \dot{\phi}_n^* - \sum_{m}  \Omega_{nm} \phi_{m}^*, \nonumber\\
\pi_n^* &=& \dot{\phi}_n + \sum_{m} \phi_{m} \Omega_{mn} = \dot{\phi}_n - \sum_{m} \Omega_{nm} \phi_{m},
\end{eqnarray}
and the time-dependent Landau energies are
\begin{eqnarray}
\omega_n^2 (t) = |qB(t)| (2n+1) + m^2 + k_z^2. \label{lan en}
\end{eqnarray}
Note that other choice of the basis, for instance, the initial Landau states at $t = - \infty$ with $\Omega = 0$, may be used, but the transverse Hamiltonian still has off-diagonal terms, $\langle m, t_0 \vert \hat{H}_{\perp} (t) \vert n, t_0 \rangle \neq 0$ for $m = n \pm 2$.
Further note that in the second quantization, we quantize not the Landau states (\ref{lan ham}) but the field modes $\phi_{n} (t)$ and $\pi_{n} (t)$ and their complex conjugates. In fact, the Landau states (\ref{lan ham}) are used as a mathematical tool to expand the field (\ref{lan exp}). The Heisenberg equation
\begin{eqnarray}
\ddot{\phi}_n - 2 \Omega_{nm} \dot{\phi}_m + \Bigl( \omega_n^2 (t) \delta_{nm} + (\Omega^2)_{nm} - \dot{\Omega}_{nm} \Bigr) \phi_m = 0, \label{B-eq}
\end{eqnarray}
is the vector component of the KG equation \cite{kim14a}.

\section{Time-Dependent Schr\"{o}dinger Equation} \label{sec3}

The Hamiltonian (\ref{E-ham}) or (\ref{B-ham}) is an infinite sum of time-dependent oscillators. Using a compact notation, in which $\alpha, \beta$ stand for ${\bf k}, {\bf k}'$  for electric fields or $(m, k_z), (n, k'_z)$ for magnetic fields such that $\sum_{\alpha} = \int d^3 {\bf k}/(2\pi)^{3}$ or $ \sum_{\alpha} =  \sum_{n} \int d k_z/(2 \pi)$, and introducing the extended phase space variables
\begin{eqnarray}
Z_{\alpha} = \begin{pmatrix} \pi_{\alpha}\\ \phi^*_{\alpha} \end{pmatrix}, \quad Z^*_{\alpha} = \begin{pmatrix} \pi^*_{\alpha}\\ \phi_{\alpha} \end{pmatrix}, \label{phase}
\end{eqnarray}
the Hamiltonian can be rewritten in the quadratic form
\begin{eqnarray}
H (t) = \sum_{\alpha \beta} Z^{\dagger}_{\alpha} {\bf H}_{\alpha \beta} (t) Z_{\beta}, \label{q-ham}
\end{eqnarray}
where the Hamiltonian matrix is
\begin{eqnarray}
 {\bf H}_{\alpha \beta} (t) = \begin{pmatrix} \delta_{\alpha \beta} & 0\\ 0 & \omega_{\alpha}^2 \delta_{\alpha \beta} \end{pmatrix} \label{E-hmat}
\end{eqnarray}
for the electric field, and it is
\begin{eqnarray}
{\bf H}_{\alpha \beta} (t) = \begin{pmatrix} \delta_{\alpha \beta} & \Omega_{\alpha \beta} \\ -\Omega_{\alpha \beta} & \omega_{\alpha}^2 \delta_{\alpha \beta}\end{pmatrix} \label{B-hmat}
\end{eqnarray}
for the magnetic field. Both matrices (\ref{E-hmat}) and (\ref{B-hmat}) are Hermitian, ${\bf H}^{\dagger} = {\bf H}$, and also make the Hamiltonian (\ref{q-ham}) Hermitian.

The quantization of the Hamiltonian (\ref{q-ham}) follows from the commutation relations
\begin{eqnarray}
[\hat{\phi}_{\alpha}, \hat{\pi}_{\beta} ] = i \delta_{\alpha \beta}, \quad [\hat{\phi}^*_{\alpha}, \hat{\pi}^*_{\beta} ] = i \delta_{\alpha \beta},
\end{eqnarray}
and all other commutators vanish. In other words, the commutation relations hold in the extended phase space such that
\begin{eqnarray}
[\hat{Z}_{\alpha}, \hat{Z}^*_{\beta} ] = \begin{pmatrix} 0& -i \delta_{\alpha \beta} \\ i \delta_{\alpha \beta} & 0 \end{pmatrix} , \quad [\hat{Z}_{\alpha}, \hat{Z}_{\beta} ] = [\hat{Z}^*_{\alpha}, \hat{Z}^*_{\beta} ] = \begin{pmatrix} 0& 0 \\ 0 & 0 \end{pmatrix}.
\end{eqnarray}
Then, the quantum evolution is governed by the time-dependent Schr\"{o}dinger equation
\begin{eqnarray}
i \frac{\partial}{\partial t} \vert \Psi (t) \rangle = \sum_{\alpha \beta}  \hat{Z}^{\dagger}_{\alpha} {\bf H}_{\alpha \beta} (t) \hat{Z}_{\beta} \vert \Psi (t) \rangle. \label{osc eq}
\end{eqnarray}
In order to find the quantum states for Eq. (\ref{osc eq}), we employ the quantum invariant method, which satisfies the Liouville-von Neumann equation
\begin{eqnarray}
i \frac{\partial \hat{I} (t)}{\partial t} + [\hat{I}(t), \hat{H} (t) ] = 0. \label{ln eq}
\end{eqnarray}
Then, the exact solution to Eq. (\ref{osc eq}) is given by an eigenstate of the quantum invariant up to a time-dependent phase factor \cite{lewis-riesenfeld69}
\begin{eqnarray}
\vert \Psi (t) \rangle = \sum_{\lambda} C_{\lambda} e^{- i \int^t dt' \langle \lambda, t' \vert \hat{H} (t')
- i \partial/ \partial t' \vert \lambda, t' \rangle} \vert \lambda, t \rangle, \label{lr sol}
\end{eqnarray}
where $C_{\lambda}$ is a constant, and $\lambda$ is a constant eigenvalue corresponding to the eigenvalue problem of the invariant operator
\begin{eqnarray}
\hat{I} (t)  \vert \lambda, t \rangle = \lambda  \vert \lambda, t \rangle.
\end{eqnarray}
Considering the symmetric form of the Hamiltonian (\ref{q-ham}) from two fields $\phi$ and $\phi^*$ and their conjugate momenta, we may search for the quantum invariants in the extended phase space (\ref{phase}) of the form
\begin{eqnarray}
\hat{I} (t) = (U(t), V(t)) \frac{\hat{Z}+ \hat{Z}^*}{\sqrt{2}}, \label{q-inv}
\end{eqnarray}
where $U (t)$ and $V (t)$ are matrix-valued functions, carrying indices of $\alpha, \beta$, and are determined by Eq. (\ref{ln eq}).

\section{Invariant Operators in Time-Dependent Electric Fields} \label{sec4}

In the case of time-dependent electric fields, the Hamiltonian (\ref{E-ham}) is a system of decoupled, time-dependent oscillators, for which we may use a pair of invariants for each oscillator as the time-dependent annihilation and the creation operators, and construct the Fock space of number states \cite{MMT70,kim-kim99,kim-lee00,kim-page01,kim-page13}.
From Eq. (\ref{q-inv}), we introduce the first class invariant for each $\alpha$
\begin{eqnarray}
\hat{A}_{\alpha} (t) = \frac{i}{\sqrt{2}} \Bigl[ \varphi^*_{\alpha} (\hat{\pi}^*_{\alpha} + \hat{\pi}_{\alpha}) - \dot{\varphi}^*_{\alpha} (\hat{\phi}^*_{\alpha} + \hat{\phi}_{\alpha}) \Bigr], \label{E-in1}
\end{eqnarray}
and the second class invariant
\begin{eqnarray}
\hat{A}^{\dagger}_{\alpha} (t) = - \frac{i}{\sqrt{2}} \Bigl[ \varphi_{\alpha} (\hat{\pi}^*_{\alpha} + \hat{\pi}_{\alpha}) - \dot{\varphi}_{\alpha} (\hat{\phi}^*_{\alpha} + \hat{\phi}_{\alpha}) \Bigr]. \label{E-in2}
\end{eqnarray}
Here, $\varphi_{\alpha} (t)$ is a complex solution to the mode equation (\ref{E-eq}) and satisfies the Wronskian condition from the quantization rule
\begin{eqnarray}
\varphi_{\alpha} (t)  \dot{\varphi}^*_{\alpha} (t) - \varphi^*_{\alpha} (t) \dot{\varphi}_{\alpha} (t) = i. \label{wr con}
\end{eqnarray}
Each complex solution $\varphi_{\alpha} (t)$ determines a pair of the invariants (\ref{E-in1}) and (\ref{E-in2}), which is analogous to a Fock basis in the standard Bogoliubov transformation. However, under the assumption of ${\bf E}_{\parallel} (-\infty) = 0$ and ${\bf B} (- \infty) = {\bf B}_0$, the complex solution can be uniquely chosen such that in the electric field it coincides with the Minkowski positive frequency solution, while in the magnetic field it coincides with the standard Landau states at $t = - \infty$. In the generic, time-dependent electromagnetic field without any asymptotic limit, the quantum invariants cannot be uniquely selected, but one may employ other principle, such as the minimum uncertainty, to choose the complex solution \cite{kim-kim99}.

Note that the quantum invariants (\ref{E-in1}) and (\ref{E-in2})
\begin{eqnarray}
\hat{A}_{\alpha} (t) = \frac{1}{\sqrt{2}} \bigl( \hat{a}_{\alpha} (t) + \hat{b}_{\alpha} (t) \bigr), \nonumber\\
\hat{A}^{\dagger}_{\alpha} (t) = \frac{1}{\sqrt{2}} \bigl(\hat{a}^{\dagger}_{\alpha} (t) + \hat{b}^{\dagger}_{\alpha} (t) \bigr), \label{E-inv}
\end{eqnarray}
can be expressed in terms of the annihilation and the creation operators for particles
\begin{eqnarray}
\hat{a}_{\alpha} (t) &=& i \bigl( \varphi_{\alpha}^{*} (t) \hat{\pi}_{\alpha}^{*}  - \dot{\varphi}_{\alpha}^{*} (t) \hat{\phi}_{\alpha} \bigr), \nonumber\\
\hat{a}_{\alpha}^{\dagger} (t) &=& -i \bigl( \varphi_{\alpha} (t) \hat{\pi}_{\alpha}  - \dot{\varphi}_{\alpha} (t) \hat{\phi}_{\alpha}^{*} \bigr), \label{pa op}
\end{eqnarray}
and those for antiparticles
\begin{eqnarray}
\hat{b}_{\alpha} (t) &=& i \bigl( \varphi_{\alpha}^{*}(t) \hat{\pi}_{\alpha}  - \dot{\varphi}_{\alpha}^{*} (t) \hat{\phi}_{\alpha}^{*} \bigr),
 \nonumber\\
\hat{b}_{\alpha}^{\dagger} (t) &=& - i \bigl( \varphi_{\alpha} (t) \hat{\pi}_{\alpha}^{*}  - \dot{\varphi}_{\alpha} (t) \hat{\phi}_{\alpha} \bigr). \label{an op}
\end{eqnarray}
The commutators hold
\begin{eqnarray}
[\hat{A}_{\alpha} (t), \hat{A}^{\dagger}_{\beta} (t) ] = [\hat{a}_{\alpha} (t), \hat{a}^{\dagger}_{\beta} (t) ] =  [\hat{b}_{\alpha} (t), \hat{b}^{\dagger}_{\beta}  (t)]= \delta_{\alpha \beta}, \label{com}
\end{eqnarray}
while other commutators vanish. Thus, $\hat{A}_{\alpha} (t)$ annihilates one particle-antiparticle pair, while $\hat{A}^{\dagger}_{\alpha} (t)$ creates the pair with the same quantum $\alpha$, and the field has the canonical representation
\begin{eqnarray}
\hat{\phi} (t, {\bf x}) = \sum_{\alpha} \Bigl[ \varphi_{\alpha} (t) \hat{a}_{\alpha} (t) + \varphi_{\alpha}^{*} (t) \hat{b}_{\alpha}^{\dagger} (t) \Bigr].
\end{eqnarray}

Using the commutation relations (\ref{com}), we may construct the time-dependent particle-antiparticle states
\begin{eqnarray}
\vert k_{\alpha}, \bar{l}_{\beta}, t \rangle = \frac{(\hat{a}_{\alpha}^{\dagger} (t))^{k_{\alpha}}}{\sqrt{k_{\alpha}!}}  \frac{(\hat{b}_{\beta}^{\dagger}(t))^{l_{\beta}}}{\sqrt{l_{\beta}!}} \vert 0_{\alpha}, \bar{0}_{\beta}, t \rangle,
\end{eqnarray}
where bars denote antiparticles, and the vacuum state is given by
\begin{eqnarray}
\hat{a}_{\alpha} (t) \vert 0_{\alpha}, \bar{0}_{\beta}, t \rangle = \hat{b}_{\beta} (t) \vert 0_{\alpha}, \bar{0}_{\beta}, t \rangle = 0.
\end{eqnarray}
These states are orthonormal among themselves at the equal time
\begin{eqnarray}
\langle m_{\alpha}, \bar{n}_{\beta}, t \vert k_{\gamma}, \bar{l}_{\delta}, t \rangle = \delta_{mk} \delta_{\alpha \gamma} \delta_{nl} \delta_{\beta \delta}.
\end{eqnarray}
However, each multiple particle-antiparticle state is not a solution to Eq. (\ref{osc eq}), because
$\hat{a}_{\alpha} (t)$ and $\hat{b}_{\alpha} (t)$ cannot separately become quantum invariants for Eq. (\ref{ln eq}). Instead, the particle-antiparticle pair constitutes the quantum invariants $\hat{A}_{\alpha} (t)$ and $\hat{A}^{\dagger}_{\alpha} (t)$. Hence, the vacuum state for a given $\alpha$ should be annihilated by the zero particle-antiparticle operator
\begin{eqnarray}
\hat{A}_{\alpha} (t) \vert 0_{\alpha}, t \rangle = 0, \label{zer st}
\end{eqnarray}
and at the same time have the zero particle and antiparticle number, respectively,
\begin{eqnarray}
\langle 0_{\alpha}, t \vert \hat{a}^{\dagger}_{\alpha} (t)  \hat{a}_{\alpha} (t) \vert 0_{\alpha}, t \rangle = \langle 0_{\alpha}, t \vert \hat{b}^{\dagger}_{\alpha} (t)  \hat{b}_{\alpha} (t) \vert 0_{\alpha}, t \rangle = 0. \label{zer pa}
\end{eqnarray}
The zero particle and antiparticle content (\ref{zer pa}) excludes another null state (\ref{zer st}) with one particle or antiparticle
\begin{eqnarray}
\vert 1_{\alpha}, t \rangle_{\rm spurious}
= \sqrt{\frac{1}{2}} \Bigl(\vert 1_{\alpha}, \bar{0}_{\alpha}, t \rangle - \vert 0_{\alpha}, \bar{1}_{\alpha}, t \rangle\Bigr).
\end{eqnarray}
Indeed, $\vert 0_{\alpha}, t \rangle := \vert 0_{\alpha}, \bar{0}_{\alpha},t \rangle$ is the unique state satisfying Eqs. (\ref{zer st}) and (\ref{zer pa}) and constructs the time-dependent vacuum state for the field
\begin{eqnarray}
\vert 0, t \rangle = \prod_{\alpha} \vert 0_{\alpha}, t \rangle. \label{vac st}
\end{eqnarray}
The first excited state is obtained by acting $\hat{A}^{\dagger}_{\alpha} (t)$ on the vacuum
\begin{eqnarray}
\vert 1_{\alpha}, t \rangle := \hat{A}^{\dagger}_{\alpha} (t) \vert 0, t \rangle
= \sqrt{\frac{1}{2}} \Bigl(\vert 1_{\alpha}, \bar{0}_{\alpha}, t \rangle + \vert 0_{\alpha}, \bar{1}_{\alpha}, t \rangle\Bigr). \label{1-p-a}
\end{eqnarray}
The one particle-antiparticle state (\ref{1-p-a}) is the superposition of a particle with ${\bf k}$ and an antiparticle
with $- {\bf k}$ from Eq. (\ref{fourier}) but is not $\vert 1_{\alpha}, \bar{1}_{\alpha}, t \rangle$, as expected naively. Similarly,
acting $n$-times $\hat{A}^{\dagger}_{\alpha} (t)$ excites $n$ particle-antiparticle state for the quantum $\alpha$
\begin{eqnarray}
\vert n_{\alpha}, t \rangle = \frac{(\hat{A}^{\dagger}_{\alpha} (t))^{n_{\alpha}} }{\sqrt{n_{\alpha}!}} \vert 0, t \rangle
= \sum_{k_{\alpha}=0}^{n_{\alpha}}  \sqrt{\frac{n_{\alpha}!}{2^{\alpha} (n_{\alpha} - k_{\alpha})! k_{\alpha}!}}
\vert n_{\alpha} - k_{\alpha}, \bar{k}_{\alpha}, t \rangle, \label{n-p-a}
\end{eqnarray}
in which the statistical weight comes from the number of ways for arranging indistinguishable $n_{\alpha}$ particle-antiparticles. Any general particle-antiparticle state is a linear combination of (\ref{n-p-a}).

It is one of the merits of the invariant operators that the exact quantum state
can also be given by the eigenstates of another invariant $\hat{N}_{\alpha} (t) = \hat{A}_{\alpha}^{\dagger} (t) \hat{A}_{\alpha} (t)$ \cite{kim-page01}. Hence, the total number of initial particle-antiparticle with the quantum $\alpha$ contained in the vacuum state at $t$ is the same as that of particle-antiparticle at $t$ contained in the initial vacuum state at $t_0$:
\begin{eqnarray}
{\cal N}_{\alpha} (t) = \langle 0, t \vert \hat{A}_{\alpha}^{\dagger} (t_0) \hat{A}_{\alpha} (t_0) \vert 0, t \rangle = \langle 0, t_0 \vert \hat{A}_{\alpha}^{\dagger} (t) \hat{A}_{\alpha} (t) \vert 0, t_0 \rangle =  |\dot{\varphi}_{\alpha} (t)|^2 |\varphi_{\alpha} (t_0)|^2 + |\varphi_{\alpha} (t)|^2|\dot{\varphi}_{\alpha} (t_0)|^2.
\end{eqnarray}
Here, $\hat{A}_{\alpha} (t_0)$ and $\hat{A}_{\alpha}^{\dagger} (t_0)$ are the operators (\ref{pa op}) and (\ref{an op}) evaluated at $t_0$. Note that $\hat{A}_{\alpha}^{\dagger} (t) \hat{A}_{\alpha} (t)$,
$(\hat{A}_{\alpha}^{\dagger} (t))^2$ and $(\hat{A}_{\alpha} (t))^2$ form an S(1,1) algebra, which leads to the quantum master equation \cite{kim-schubert11}, and to the evolution operator and therefrom the pair-production rate \cite{KLR12}. Furthermore, the one-loop effective action is obtained from the scattering matrix between the out-vacuum and the in-vacuum under the action of the electric field modulo the scattering matrix in the absence of the field \cite{KLY08}:
\begin{eqnarray}
e^{i \int  d^4x {\cal L}_{\rm eff} (E)} = \prod_{\alpha, T = \infty} \frac{\langle 0_{\alpha}, T/2 \vert 0_{\alpha}, -T/2 \rangle (E) }{\langle 0_{\alpha}, T/2 \vert 0_{\alpha}, -T/2 \rangle (E=0)}.
\end{eqnarray}

\section{Invariant Operators in Time-Dependent Magnetic Fields} \label{sec5}

The charged scalar field in a time-dependent magnetic field along the fixed direction has the Hamiltonian (\ref{B-ham}) or (\ref{q-ham}), which may be written in the form
\begin{eqnarray}
H_{\epsilon} (t) = H_{\rm D} (t) + \epsilon H_{\rm O} (t). \label{B-tham}
\end{eqnarray}
Here, $\epsilon$ is an expansion parameter for the perturbation theory and will be set $\epsilon =1$ in the end of calculations, and
\begin{eqnarray}
H_{\rm D} = \sum_{\alpha} \pi^*_{\alpha} \pi_{\alpha} + \omega^2_{\alpha} (t) \phi^*_{\alpha} \phi_{\alpha} \label{diag}
\end{eqnarray}
is the diagonal Hamiltonian, while
\begin{eqnarray}
H_{\rm O} = \sum_{\alpha \beta} \pi^*_{\alpha} \Omega_{\alpha \beta} \phi^*_{\beta} + \pi_{\alpha} \Omega_{\alpha \beta} \phi_{\beta} \label{off}
\end{eqnarray}
is the off-diagonal Hamiltonian.
The Hamiltonian (\ref{B-tham}) is an infinite system of coupled oscillators due to the off-diagonal Hamiltonian (\ref{off}), which induces continuous transitions among Landau levels as shown in Eq. (\ref{ll tr}). In the non-relativistic theory, a charged particle in time-dependent magnetic fields
similarly has a finite number of coupled oscillators, to which the quantum invariant method has been applied \cite{lewis-riesenfeld69,fiore-gouba11} (for coupled, time-dependent oscillators, see  also Refs. \cite{leach77,ji-hong98,dodonov00,lee01}). The infinite degrees of freedom for the charged field necessarily involve renormalizing physical quantities such as the vacuum energy and the charge, in strong contrast to finite degrees of the freedom for the particle.

The quantum motion of the charged field may be classified according to the relative ratio $||H_{\rm O}|| / ||H_{\rm D}||$ for some appropriate measure. For instance, a measure has been introduced that the integrated rate of the change of Landau levels to the dynamical phase for any time interval \cite{kim14a}
\begin{eqnarray}
{\cal R}_{\alpha} =  \frac{\int_{t_a}^{t_b} dt' |\Omega_{\alpha \beta} (t')|}{ \int_{t_a}^{t_b} dt' \omega_{\alpha} (t')},
\end{eqnarray}
classifies the quantum motions (states) into the adiabatic change, the sudden change, and the nonadiabatic change. Now, these motions are classified by the relative ratio:
(i) the adiabatic change when $||H_{\rm D}|| \gg  ||H_{\rm O}||$, static fields being an extreme limit of $||H_{\rm O}|| =0$, (ii) the sudden change when $||H_{\rm D}|| \ll ||H_{\rm O}||$, in which Landau levels change more rapidly than Landau energies, and (iii) the nonadiabatic change when $||H_{\rm D}|| \approx ||H_{\rm O}||$, in which the change of Landau states is comparable to the dynamical phase of Landau energies. The exact quantum motion may not be found for general time-dependent magnetic fields, except for some limiting cases such as a constant magnetic field and a suddenly changing field from one constant value to another. Hence, we propose another scheme which applies a perturbation theory to the Hamiltonian (\ref{B-tham}), by treating $H_{\rm D}$ as the unperturbed Hamiltonian and $H_{\rm O}$ as the perturbed one when $||H_{\rm D}|| \gg ||H_{\rm O}||$ or vice versa.

From Eq. (\ref{ln eq}), the pair of quantum invariants (\ref{q-inv}) in the extended phase space (\ref{phase})
\begin{eqnarray}
\hat{I}_{(\pm)} (t) = \frac{1}{\sqrt{2}} \Bigl[ U_{_{(\pm)} \alpha \beta} (\hat{\pi}^*_{\beta} + \hat{\pi}_{\beta}) + V_{(\pm) \alpha \beta} (\hat{\phi}^*_{\beta} + \hat{\phi}_{\beta}) \Bigr], \label{q-inv2}
\end{eqnarray}
may be found by solving the matrix-valued differential equations
\begin{eqnarray}
\dot{U}_{(\pm)} + U_{(\pm)}  \Omega + V_{(\pm)} = 0, \quad \dot{V}_{(\pm)} - U_{(\pm)} {\bf \omega}^2 + V_{(\pm)} \Omega = 0. \label{uv eq}
\end{eqnarray}
Here, the positive (negative) sign denotes the positive (negative) frequency solution.
The two equations (\ref{uv eq}) are equal to the transpose of the mode equation in the vector form (\ref{B-eq}),
\begin{eqnarray}
\ddot{U}_{(\pm)} + 2 \dot{U}_{(\pm)} \Omega + U_{(\pm)} \bigl({\bf \omega}^2 + \Omega^2 + \dot{\Omega} \bigr) = 0. \label{B-teq}
\end{eqnarray}
However, the merit of the quantum invariant method does not lie in solving the classical field equation, but lies in constructing diverse quantum states. In the two-component first-order formalism, the Cauchy problem has been reduced to solving Eq. (\ref{B-eq}), whose solutions may not be known for generic magnetic fields, and requires a perturbation method. In a similar manner, finding the exact invariants (\ref{q-inv2}) is equivalent to solving Eq. (\ref{B-teq}) or (\ref{B-eq}). Therefore, we need a systematic method, such as perturbation theory, in order to find the invariants or directly to solve the time-dependent Schr\"{o}dinger equation.

\subsection{Adiabatic Change} \label{sec5-1}

The field $\phi$ is complex due to the longitudinal motion along the magnetic field in the four-dimensional spacetime, so we may quantize the time-dependent oscillators (\ref{diag}) as for electric fields in Sec. \ref{sec4}. Hence, the invariant operators are
\begin{eqnarray}
\hat{\cal A}_{\alpha} (t) &=& \frac{i}{\sqrt{2}} \Bigl[ \varphi^*_{\alpha} (\hat{\pi}^*_{\alpha} + \hat{\pi}_{\alpha}) - \dot{\varphi^*}_{\alpha} (\hat{\phi}^*_{\alpha} + \hat{\phi}_{\alpha}) \Bigr]
= \frac{1}{\sqrt{2}} \bigl( \hat{a}_{\alpha} (t) + \hat{b}_{\alpha} (t)  \bigr), \nonumber\\
\hat{\cal A}^{\dagger}_{\alpha} (t) &=& - \frac{i}{\sqrt{2}} \Bigl[ \varphi_{\alpha} (\hat{\pi}^*_{\alpha} + \hat{\pi}_{\alpha}) - \dot{\varphi}_{\alpha} (\hat{\phi}^*_{\alpha} + \hat{\phi}_{\alpha}) \Bigr] = \frac{1}{\sqrt{2}} \bigl( \hat{a}^{\dagger}_{\alpha} (t) + \hat{b}^{\dagger}_{\alpha} (t)  \bigr). \label{B-in}
\end{eqnarray}
Here, the auxiliary field $\varphi_{\alpha}$ for $\alpha = (n, k_z)$ is a complex solution to the mode equation
\begin{eqnarray}
\ddot{\varphi}_{\alpha} (t) + \omega^2_{\alpha} (t) \varphi_{\alpha} (t) = 0 \label{aux sol2}
\end{eqnarray}
with $\omega^2_{\alpha} = |qB| (2n+1) + m^2 + k^2_z$ and  ${\rm Wr} [\varphi_{\alpha}, \varphi^*_{\alpha} ] = i$.
Then, the diagonal Hamiltonian becomes
\begin{eqnarray}
H_{\rm D} = \sum_{\alpha} \Bigl[ \bigl( \dot{\varphi}^*_{\alpha} \dot{\varphi}_{\alpha} + \omega^2_{\alpha} \varphi^*_{\alpha} \varphi_{\alpha} \bigr) \bigl(\hat{a}^{\dagger}_{\alpha} \hat{a}_{\alpha} + \hat{b}_{\alpha}\hat{b}^{\dagger}_{\alpha} \bigr) + \bigl( \dot{\varphi}^{*2}_{\alpha} + \omega^2_{\alpha} \varphi^{*2}_{\alpha} \bigr) \hat{a}^{\dagger}_{\alpha} \hat{b}^{\dagger}_{\alpha} + \bigl( \dot{\varphi}^{2}_{\alpha} + \omega^2_{\alpha} \varphi^{2}_{\alpha} \bigr) \hat{a}_{\alpha} \hat{b}_{\alpha} \Bigr], \label{diag2}
\end{eqnarray}
and the off-diagonal Hamiltonian takes the form
\begin{eqnarray}
H_{\rm O} = \sum_{\alpha \beta} \Bigl[ \bigl(  \dot{\varphi}^*_{\alpha} \varphi_{\beta} -  \varphi^*_{\alpha} \dot{\varphi}_{\beta} \bigr) \Omega_{\alpha \beta} \bigl( \hat{a}^{\dagger}_{\alpha} \hat{a}_{\alpha} + \hat{b}_{\alpha}\hat{b}^{\dagger}_{\alpha} \bigr)
+ \bigl( \dot{\varphi}^*_{\alpha} \varphi^*_{\beta} - \varphi^*_{\alpha} \dot{\varphi}^*_{\beta} \bigr) \Omega_{\alpha \beta} \hat{a}^{\dagger}_{\alpha} \hat{b}^{\dagger}_{\beta}
+ \bigl( \dot{\varphi}_{\alpha} \varphi_{\beta} - \varphi_{\alpha} \dot{\varphi}_{\beta} \bigr) \Omega_{\alpha \beta} \hat{a}_{\alpha} \hat{b}_{\beta} \Bigr]. \label{off2}
\end{eqnarray}
The invariant operators (\ref{B-in}) satisfy Eq. (\ref{ln eq}) for the diagonal Hamiltonian (\ref{diag2}) when $\varphi_{\alpha}$ is the solution to Eq. (\ref{aux sol2}).

In the adiabatic change, we find the quantum states for the unperturbed Hamiltonian (\ref{diag2}) with the aid of the quantum invariant method and then improve the unperturbed states by the Hamiltonian (\ref{off2}). The unperturbed states consist of the time-dependent vacuum state
\begin{eqnarray}
\hat{a}_{\alpha} (t) \vert 0_{\alpha}, t \rangle_{(0)} = \hat{b}_{\alpha} (t) \vert 0_{\alpha}, t \rangle_{(0)} =0, \label{lan qst}
\end{eqnarray}
and the excited states
\begin{eqnarray}
\vert n_{\alpha}, t \rangle_{(0)} = \frac{(\hat{\cal A}_{\alpha} (t))^{n_{\alpha}}}{\sqrt{n_{\alpha}!}} \vert 0_{\alpha}, t \rangle_{(0)}.
\end{eqnarray}
For each $\alpha$ the time-dependent vacuum state (\ref{lan qst}) leads to the Landau state and thereby
the time-dependent vacuum state for the field
\begin{eqnarray}
\vert 0, t \rangle_{(0)} = \prod_{\alpha} \vert 0_{\alpha}, t \rangle_{(0)}. \label{B-vac}
\end{eqnarray}
Furthermore, the general excited states are \cite{kim-khanna00}
\begin{eqnarray}
\vert {\cal N}, t \rangle_{(0)} = \prod_{{\cal N}} \frac{\hat{\cal A}^{\dagger {\cal N}} (t)}{\sqrt{ \{ {\cal N} \}!}} \vert 0, t \rangle_{(0)}
\end{eqnarray}
where $\{ {\cal N} \} = (n_0, \cdots, n_{\alpha}, \cdots)$, $\hat{\cal A}^{\dagger {\cal N}} = \hat{\cal A}_{0}^{\dagger n_0} \cdots \hat{\cal A}_{\alpha}^{\dagger n_{\alpha}} \cdots$, and
$ \{{\cal N} \}! = \prod_{\alpha} n_{\alpha}!$.

The quantum state of the field in a constant magnetic field $B_0$, though looks trivial because of  $\Omega_{\alpha \beta} = 0$, requires a careful understanding in contrast to that of a charged particle in the non-relativistic theory. Each Landau level has the solution
\begin{eqnarray}
\varphi_{\alpha} (t) = \frac{ e^{- i \omega_{\alpha} t}}{\sqrt{2 \omega_{\alpha}}}
\end{eqnarray}
with the Landau energy (\ref{lan en}).
Then, the exact vacuum state (\ref{lr sol}) for the time-dependent Schr\"{o}dinger equation is given by
\begin{eqnarray}
\vert 0, t \rangle = \prod_{\alpha} e^{- i \omega_{\alpha} t} \vert 0_{\alpha} \rangle.
\end{eqnarray}
Note that $\prod_{\alpha} \vert 0_{\alpha} \rangle$ is the vacuum state associated with $\Phi_{n} ({\bf x}_{\perp}, t)$ in Eq. (\ref{lan exp}), in which each Landau state is equally occupied. The scattering matrix between the out-vacuum and the in-vacuum gives the one-loop effective action
\begin{eqnarray}
\langle 0, \frac{T}{2} \vert 0, - \frac{T}{2} \rangle = e^{i T V_{\perp}  \sum_{\alpha} \omega_{\alpha} (B_0) } = e^{i T V {\cal L}_{\rm eff} (B_0)},
\end{eqnarray}
where $V$ and $V_{\perp}$ are the three- and two-dimensional volumes of the problem. In fact, the summation of all the Landau energies equals to the one-loop effective action after the renormalization of the vacuum energy and the charge \cite{greiner-reinhardt}.

For adiabatically changing magnetic fields, the scattering matrix between the out-vacuum and the in-vacuum provides the vacuum persistence amplitude
\begin{eqnarray}
{}_{(0)}\langle 0, \infty \vert 0, - \infty \rangle_{(0)} = e^{i \int d^4 x {\cal L}_{\rm eff}}. \label{vacum per}
\end{eqnarray}
The non-unimodular amplitude $|{}_{(0)}\langle 0, \infty \vert 0, - \infty \rangle_{(0)} |^2 < 1$ implies that the initial vacuum state is not stable against the pair production and that the effective action obtains an imaginary part. The instability can be expected from the induced electric field ${\bf E} = - \partial {\bf A}/ \partial t$. However, the adiabatic vacuum state (\ref{B-vac}) is no longer a good approximation for rapidly changing magnetic fields because $||H_{\rm D}|| \ll ||H_{\rm O}||$.

\subsection{Perturbation beyond Adiabatic Change} \label{sec5-2}

In the time-dependent perturbation theory \cite{messiah}, the time-dependent vacuum state (\ref{B-vac}) does not lead to any quantum correction because
\begin{eqnarray}
{}_{(0)}\langle 0, t \vert \hat{H}_{\rm O} \vert 0, t \rangle_{(0)} = 0.
\end{eqnarray}
It would be interesting to compare the Hamiltonian (\ref{B-tham}) with the Fourier-decomposed Hamiltonian for an interacting scalar field, for instance, the $\Phi^4$-theory \cite{kim-khanna00}.  We may look for the improved state
\begin{eqnarray}
\vert {\cal N}, t \rangle = \hat{U} (\hat{\cal A}^{\dagger} (t), \hat{\cal A} (t), t) \vert {\cal N}, t \rangle_{(0)},
\end{eqnarray}
and solve the time-dependent Schr\"{o}dinger equation
\begin{eqnarray}
i \frac{\partial}{\partial t} \hat{U} (t)  = \hat{H}_{\rm O} (t) \hat{U} (t). \label{ev op}
\end{eqnarray}
It is the property of the invariant operators that the perturbation $\hat{H}_{\rm O} (t)$ is invariant in the interaction-like picture \cite{kim-khanna00}.
Thus, the evolution operator for Eq. (\ref{ev op}) has the formal solution given by the time-ordered integral
\begin{eqnarray}
\hat{U} (t) = {\rm T} \exp \Bigl[ -i \int_{t_0}^t dt' \hat{H}_{\rm O} (t') \Bigr],
\end{eqnarray}
and the leading approximation is
\begin{eqnarray}
\hat{U}_{(1)} (t) =  e^{-i \int_{t_0}^t dt' \hat{H}_{\rm O} (t')}. \label{imp st}
\end{eqnarray}
Then, the vacuum persistence amplitude between the in-vacuum and the out-vacuum, which is improved by Eq. (\ref{imp st}) when the magnetic field changes slowly during a large period $T$ from one value at $- T/2$ to another at $T/2$, is given by
\begin{eqnarray}
e^{ i V \int dt' {\cal L}_{\rm eff}} = {}_{(0)}\langle 0, \frac{T}{2} \vert e^{i \int_{-T/2}^{T/2} dt' \hat{H}_{\rm O} (t')} \vert 0, - \frac{T}{2} \rangle_{(0)}.
\end{eqnarray}

\subsection{Electric Field Parallel to Magnetic Field} \label{sec5-3}

When a time-dependent electric field parallel to the magnetic field is present, the vector potential (\ref{vec pot}) may be used. Then, the Hamiltonian after the spectral decomposition (\ref{lan exp}) is given by
\begin{eqnarray}
H(t) = \sum_{\alpha} \bigl( \pi_{\alpha}^* \pi_{\alpha} + \omega_{\alpha}^2 \phi_{\alpha}^* \phi_{\alpha} \bigr) + \sum_{\alpha \beta} \bigl( \pi_{\alpha} \Omega_{\alpha \beta} \phi_{\beta} +  \pi_{\alpha}^* \Omega_{\alpha \beta}  \phi_{\beta}^* \bigr), \label{EB-ham}
\end{eqnarray}
where $\alpha = (n, k_z)$ and the time-dependent Landau energies are
\begin{eqnarray}
\omega_{\alpha}^2 (t) = |qB(t)| (2n+1) + m^2 + (k_z - qA_z (t))^2.
\end{eqnarray}
The nonstationary nature of quantum states does not essentially change as far as the magnetic field is time-dependent or an electric field is present. Note that the quantum motion becomes nonstationary even for a constant magnetic field due to the electric field, which is the reasoning behind the pair production due to the electric field. In the case of the adiabatic change of the magnetic field, we may use the quantum states in Sec. \ref{sec5-1} and the improved states in Sec. \ref{sec5-2}, which will not be repeated here.

\section{Conclusion} \label{sec6}

In scalar QED, we have formulated the second quantized scalar field in a homogeneous, time-dependent electromagnetic field, in which the quantum law is the functional Schr\"{o}dinger equation with a time-dependent Hamiltonian from the field action. The Hamiltonian obtained from the spectral method is quadratic in position and momentum operators, which is an infinite sum of decoupled, time-dependent oscillators in the time-dependent electric field, but is another infinite sum of coupled, time-dependent oscillators in the time-dependent magnetic field due to continuous transitions of Landau levels. The quantum law is thus the time-dependent Schr\"{o}dinger equation for an infinite system of time-dependent oscillators.

We have then employed the quantum invariant method to find the quantum states for the charged scalar field, which makes use of quantum invariants that satisfy the Liouville-von Neumann equation with respect to the time-dependent Hamiltonian \cite{lewis-riesenfeld69}. We have further used the time-dependent annihilation and the creation operators, quantum invariants, and constructed the Fock space of all the excited states \cite{MMT70,kim-kim99,kim-lee00,kim-page01}. In the case of time-dependent electric fields, the quantum law reduces to finding the wave functions for each time-dependent oscillator, while in the case of time-dependent magnetic fields, the task is equivalent to solving the Schr\"{o}dinger equation for coupled oscillators with time-dependent frequencies and couplings among different oscillators. We have sought for the quantum invariants for the scalar field in the time-dependent magnetic fields.

The quantum invariants directly lead to the exact quantum states for decoupled or coupled, time-dependent oscillators up to time-dependent phase factors. In particular, the time-dependent annihilation and the creation operators construct not only the Fock of all the excited states but also generalized correlated states and thermal states. The two-dimensional Landau states have been studied in time-dependent magnetic fields in the non-relativistic theory \cite{lewis-riesenfeld69,fiore-gouba11}. The second quantized scalar field is a relativistic theory and is equivalent to an infinite number of time-dependent oscillators as explored in this paper. We have advanced a perturbation method based on the quantum invariants for the time-dependent magnetic fields because the solutions for the classical mode equations are not known in general. In particular, when the magnetic fields change adiabatically in the sense that the rate of the change of the Landau levels is smaller than the corresponding dynamical phases, the off-diagonal Hamiltonian responsible for the transitions among the time-dependent Landau levels could be treated as perturbation to the adiabatic, diagonal Hamiltonian from the Landau levels. The perturbation improves the time-dependent vacuum state of Landau levels.

The quantum state or wave functional for the field in QED always involves some infinite quantity which should be regulated away through the renormalization of physical quantities, such as the vacuum energy and the charge of the particle. For instance, the one-loop effective action via the scattering matrix between the out-vacuum at the remote future and the in-vacuum in the remote past is given by the sum over all the quantum numbers, the three-momenta or the Landau levels and the longitudinal momenta. The renormalized action follows from an appropriate regularization scheme; for instance, in a time-dependent Sauter-type electric field, the complex mode solution (\ref{E-eq}) is known, the time-dependent vacuum state can be constructed from the invariants, and the renormalization of the action is essentially the same as Ref. \cite{KLY08}. The renormalization problem also occurs for a scalar field in an expanding universe \cite{ringwald87}. The renormalization problem in the second quantized scalar field may be studied on a case-by-case basis, which goes beyond the scope of this paper.

There are some interesting issues not handled in this paper but requiring a further study. The external electromagnetic fields that have both temporal and spatial distribution require a more advanced spectral method, which goes beyond the scope of this paper. Another interesting problem is a time-dependent magnetic field whose direction rotates around a fixed direction \cite{dipiazza-calucci02}, which may be used as a model for highly magnetized neutron stars. It would be also interesting to compare the second quantized scalar field in a rotating electric field with the recent Wigner function formalism \cite{blinne-gies13}.
Still another problem is the second quantized formulation of charged spin-1/2 fermions in the time-dependent magnetic fields, which will be addressed in a future publication.

\acknowledgments

The author thanks  Hyun Kyu Lee and Yongsung Yoon for helpful discussions on Landau states and the adiabatic theorem in magnetic fields, and thanks Lee Lindblom for useful discussions on the spectral method during the AP School on Gravitation and Cosmology at Academia Sinica, Taiwan,  2014.
He also would like to thank Misao Sasaki and Takahiro Tanaka for the warm hospitality at Yukawa Institute for Theoretical Physics, Kyoto University,
Don N. Page for the warm hospitality at University of Alberta, Jeremy S. Heyl for the warm hospitality at Pacific Institute for Theoretical Physics, University of British Columbia, and Toshiki Tajima for the warm hospitality at University of California, Irvine, where parts of this paper were done, respectively. The revision of this paper was done at University of Alberta. The visits to Yukawa Institute for Theoretical Physics, University of Alberta, Pacific Institute for Theoretical Physics, and Academia Sinica were supported by the Basic Science Research Program through the National Research Foundation of Korea (NRF) funded by the Ministry of Education (NRF-2012R1A1B3002852). This work was supported by the Research Center Program of IBS (Institute for Basic Science) in Korea.

\end{document}